# A mathematical Study of Magnetohydrodynamic Casson Fluid via Special Functions with Heat and Mass Transfer embedded in Porous Plate

Kashif Ali Abro[1], Hina Saeed Shaikh[2], Ilyas Khan[3]

[1]Department of Basic Science and Related Studies, Mehran University of Engineering Technology, Jamshoro, Pakistan
[2]Federal Government College, Karachi, Pakistan
[3]Basic Engineering Sciences Department, College of Engineering Majmaah University, Saudi Arabia

Correspondence should be addressed to Kashif Ali, kashif.abro@faculty.muet.edu.pk

**Abstract**

This article is proposed to investigate the impacts of heat and mass transfer in magnetohydrodynamic casson fluid embedded in porous medium. The generalized solutions have been traced out for the temperature distribution, mass concentration and velocity profiles under the existence and non-existence of transverse magnetic field, permeability and porosity. The corresponding solutions of temperature distribution and mass concentration, velocity profiles are expressed in terms of newly defined generalized Robotnov-Hartley function, wright function and Mittage-Leffler function respectively. All the corresponding solutions fulfill necessary conditions (initial, natural and boundary conditions) as well. Caputo Fractionalized solutions have been converted for ordinary solutions by substituting $\zeta = 1$. Some similar solutions for the temperature distribution, mass concentration and velocity profiles have been particularized form generalized solutions. Owing to the rheology of problem, graphical illustrations of distinct parameters are discussed in detail by depicting figures using Mathcad software (15).

## 1. Introduction

Due to abundant applications of non-Newtonian fluids in technological development and advancement, many engineers and scientists are working on distinct investigations such as cosmetics, pharmaceuticals, chemicals, oil, gas, food and several others. Even non-Newtonian fluids are not easy to tackle in comparison with Newtonian fluids. This happens due to the non-availability of at least single constitutive equation that can give explanations of all characteristics in non-Newtonian fluids. In order to have the explanations of all characteristics of non-Newtonian models have been presented for instance, Walters-B [1], Oldroyd-B [2], Jeffrey [3], Bingham plastic [4], power law [5], Brinkman type [6], viscoplastic [7], Maxwell [8, 9], second grade [10]. In continuation, the most popular model of non-Newtonian fluid is known as casson model [11]. Casson model is used in pigment oil suspensions for the predictions of behavior of fluid flows. This model is highly configured by several researchers in distinct situations of fluid flows. Malik et al. have investigated vertical exponentially stretching cylinder for boundary layer flow of Casson fluid [12]. Venkatesan et al. analyzed stenosed narrow arteries for blood rheology for Casson fluid under mathematical study [13]. Taza

Gul et al. have perused MHD third grade fluid under assumptions of no slip boundary condition for vertical belt with thin flim flow [14]. In this paper they investigated analytical expression for energy and momentum equations by employing Adomian decomposition method. Sidra et al. analyzed two vertical plates for unsteady second grade fluid under assumptions of oscillatory boundary and conditions and magnetic field [15]. They explored the exact solutions for temperature distribution and velocity field from the set of non-linear partial differential equations. It is focused point in their paper that they emphasized the thermal effects on the vertical plates. The generalized third grade fluid for Poiseuille and Couette flows in the presence and absence of magnetohydrodynamics has been led by Rasheed and et al. [16]. They considered the flow between two parallel plates and trace out the non-linear partial differential equation using Homotopy Perturbation Method. Kashif analyzed second grade fluid in porous medium for oscillations of plate [17]. In continuation, he extended the work of [17] for fractionalized viscoelastic fluid under influences of magnetic field and expressed the general solutions in term of generalized Fox-H function [18]. Fractional differential equation has become a fundamental tool for the modeling of various physical phenomenon's, for instance seepage flow in porous media, nonlinear oscillations of earthquakes, fluid dynamic traffic models, electrochemistry, optimal control, electromagnetism, viscoelasticity and several others. The usefulness of fractional calculus in the differential equations govern the fluid problem is the replacement of time derivative of integer order with fractional derivative $\alpha, \beta$ or $\gamma$. According to authors' cognizance, fractional calculus is rarely used in heat and mass transfer analysis under influence of magnetic field embedded in porous medium. However, our aim is to investigate the impacts of heat and mass transfer in magnetohydrodynamic casson fluid embedded in porous medium. The generalized solutions have been traced out for the temperature distribution, mass concentration and velocity profiles under the existence and non-existence of transverse magnetic field and porosity. The corresponding solutions of temperature distribution and mass concentration, velocity profiles are expressed in terms of newly defined generalized Robotnov-Hartley function, wright function and Mittage-Leffler function respectively. All the corresponding solutions fulfill necessary conditions (initial, natural and boundary condition) as well. Caputo Fractionalized solutions have been converted for ordinary solutions by substituting $\zeta = 1$. Some similar solutions for the temperature distribution, mass concentration and velocity profiles have been particularized form generalized solutions. In order to have some cognizance about the behavior of fluid, the graphical illustrations for distinct parameters such as fractional and rheological parameters, prantl number, transverse magnetic field, permeability and few others parameters have been portrayed for fluid flows.

## 2. Governing Equations

Let us assume unsteady and electrically conducting flow in casson fluid with heat and mass transfer along with oscillating plate in porous medium at $y = 0$ and $y$ is the coordinate axis normal to the plate. At the beginning when $t = 0$, the temperature $T_\infty$ for the fluid and plate are at rest conditions. When $t = 0^+$, that begins in oscillations in its own plane. Simultaneously, heat and mass transfer from the plate to the fluid proportionate to the temperature. We assume the rheology of equations for incompressible and isotropic casson fluid as in [19].

$$\tau = \mu \dot{\sigma} + \tau_0, \tag{1}$$

or

$$\tau_{ij} = \{2e_{ij}(p_y/\sqrt{2\pi} + \mu_B), \ \pi_c < \pi\} = \{2e_{ij}(p_y/\sqrt{2\pi_c} + \mu_B), \ \pi > \pi_c\}, \tag{2}$$

Here, $\tau, \mu, \dot{\sigma}, \tau_0, e_{ij}, p_y, \pi, \pi_c, \mu_B$ are shear stress, dynamic viscosity, shear rate, casson yield stress, deformation rate, yield stress, product of components of deformation rate, critical value of non-Newtonian fluid, plastic dynamic viscosity respectively. Under the above specification and assumption with Boussinesq approximation under non-dimensionalization, we have following fractionalized differential equations [20]

$$\frac{P_r}{(1+N)} D_t^\zeta T(y,t) - \frac{\partial^2 T(y,t)}{\partial y^2} = 0, \tag{3}$$

$$S_c D_t^\zeta C(y,t) - \frac{\partial^2 C(y,t)}{\partial y^2} = 0, \tag{4}$$

$$D_t^\zeta V(y,t) - \left(1 + \frac{1}{\alpha}\right) \frac{\partial^2 V(y,t)}{\partial y^2} - G_r T(y,t) - G_m C(y,t) + MV(y,t) + \frac{\emptyset}{k} V(y,t) = 0. \tag{5}$$

Where, $P_r$, $N$, $S_c$, $\zeta$, $G_r$, $G_m$, $M$, $\emptyset$, $k$, $\alpha$, $T(y,t)$, $C(y,t)$, $V(y,t)$ are prandtl number, thermal radiation, schmidt number, caputo fractional parameter, thermal Grashof number, mass Grashof number, magnetic field, porosity, permeability, material parameter of casson fluid, temperature distribution, mass concentration, velocity profile respectively and $D_t^\zeta$ defined as

$$D_t^\zeta G(t) = \begin{cases} \dfrac{1}{\Gamma(1-\zeta)} \int_0^t \dfrac{g'(p)}{(t-p)^\zeta} dp, & 0 < \zeta < 1 \\ \dfrac{dG(t)}{dt}, & \zeta = 1 \end{cases} \tag{6}$$

Equation (6) is the time fractional derivative operator given by Caputo [21]. The corresponding necessary conditions are

$$T(y,0) = C(y,0) = V(y,0) = 0; \quad y \geq 0, \tag{7}$$

$$T(y,t) \to C(y,t) \to V(y,t) \to 0; \quad as \quad y \to \infty, \tag{8}$$

$$T(0,t) = t, \quad C(0,t) = 1, \quad V(0,t) = UH(t)\cos(\omega t)/\sin(\omega t). \tag{9}$$

Equations (7), (8) and (9) are initial, natural and boundary conditions respectively.

### 3. Investigation of Temperature Distribution

For perusing the solution of temperature distribution, we apply Laplace transform on fractionalized differential equation (3) under consideration of equations $(7_1)$, $(8_1)$ and $(9_1)$, we attain

$$\frac{P_r s^\zeta \bar{T}(y,s)}{(1+N)} - \frac{\partial^2 \bar{T}(y,s)}{\partial y^2} = 0, \tag{10}$$

Writing equation (10) equivalently

$$\bar{T}(y,s) = \frac{1}{s^2} e^{-y\sqrt{\frac{p_r\, s^\zeta}{(1+N)}}}, \tag{11}$$

Inverting equation (11) by Laplace transform and using the fact of inverse Laplace transform $\mathcal{L}^{-1}\left[ exp(-\lambda_2 s^{\lambda_3})s^{-\lambda_1}\right] = \Phi(\lambda_1, -\lambda_3; -\lambda_2 t^{-\lambda_3})\, t^{\lambda_1 - 1}$, $\lambda_2, \lambda_1 \geq 0$, $0 < \lambda_3 < 1$, [22, 23] we obtain

$$T(y,s) = \Phi\left[2, -\frac{\zeta}{2}; y\sqrt{\frac{p_r}{t^\zeta(1+N)}}\right] t, \tag{12}$$

Where, the $\Phi(\lambda_1, -\lambda_2; \lambda_3)$ is the wright function [22, 23] defined as

$$\sum_{\gamma=0}^{\infty} \frac{(\lambda_3)^\gamma}{\gamma!\, \Gamma(\lambda_1 - \lambda_2 \gamma)} = \Phi(\lambda_1, -\lambda_2; \lambda_3), \qquad 0 < \lambda_2 < 1, \tag{13}$$

Equation (12) fulfills the initial condition, natural condition and boundary condition $(7_1)$, $(8_1)$ and $(9_1)$ respectively.

### 4. Investigation of Mass Concentration

For exploring the general solution of Mass concentration, we apply Laplace transform on fractionalized differential equation (4) under consideration of equations $(7_2)$, $(8_2)$ and $(9_2)$, we have

$$S_c\, s^\zeta\, \bar{C}(y,s) - \frac{\partial^2 \bar{C}(y,s)}{\partial y^2} = 0, \tag{14}$$

suitable expression of equation (14) is

$$\bar{C}(y,s) = \frac{1}{s} e^{-y\sqrt{S_c\, s^\zeta}}, \tag{15}$$

In order to have the solution of mass concentration in terms of generalized wright function, we express equation (15) in series form as

$$\bar{C}(y,s) = \sum_{j=0}^{\infty} \frac{\left(-y\sqrt{S_c}\right)^j}{j!} s^{\frac{j\zeta}{2}-1}, \tag{16}$$

Inverting equation (16) by means of Laplace transform, we get

$$C(y,s) = \sum_{j=0}^{\infty} \frac{\left(-y\sqrt{S_c}\right)^j}{j!\, \Gamma\left(-\frac{j\zeta}{2}+1\right)} t^{-\frac{j\zeta}{2}}, \tag{17}$$

Expressing equation (17) in the form of generalized wright function, we get general solution of mass concentration as

$$C(y,t) = \mathbf{W}_{-\frac{\zeta}{2},1}\left(-y\sqrt{\frac{S_c}{t^\zeta}}\right), \tag{18}$$

Where the property of generalized wright function is

$$\sum_{\gamma=0}^{\infty} \frac{(\lambda_1)^\gamma}{\gamma!\,\Gamma(\lambda_2\gamma + \lambda_3)} = \mathbf{W}_{\lambda_2,\lambda_3}(\lambda_1), \tag{19}$$

Equation (18) fulfills the initial condition, natural condition and boundary condition $(7_2)$, $(8_2)$ and $(9_2)$ respectively.

## 5. Investigation of Velocity Profiles

**Case-I:** $V(0,t) = UH(t)\cos(\omega t)$

Applying Laplace transform on equation (5) having in mind the equations $(7_3)$, $(8_3)$ and $(9_3)$, we traced

$$s^\zeta \bar{V}(y,s) - \left(1 + \frac{1}{\alpha}\right)\frac{\partial^2 \bar{V}(y,s)}{\partial y^2} + M\bar{V}(y,s) + \frac{\phi}{k}\bar{V}(y,s) - G_r\bar{T}(y,s) - G_m\bar{C}(y,s) = 0. \tag{20}$$

Employing equation (11) and (15) into equation (20), we have simplified form as

$$\bar{V}(y,s) = \frac{Us\, e^{-y\sqrt{\frac{s^\zeta + \frac{\phi}{k} + M}{1+\frac{1}{\alpha}}}}}{s^2 + \omega^2} - \frac{G_r(1+N)\, e^{-y\sqrt{\frac{p_r s^\zeta}{1+N}}}}{s^2\left[\left(1+\frac{1}{\alpha}\right)p_r s^\zeta - (1+N)\left(s^\zeta + \frac{\phi}{k} + M\right)\right]} - \frac{G_m\, e^{-y\sqrt{S_c s^\zeta}}}{s\left[\left(1+\frac{1}{\alpha}\right)S_c s^\zeta - \left(s^\zeta + \frac{\phi}{k} + M\right)\right]}, \tag{21}$$

expanding equation (21) takes the form in series as

$$\bar{V}(y,s) = \frac{Us}{s^2+\omega^2} + \frac{Us}{s^2+\omega^2}\sum_{l=1}^{\infty}\left(\frac{-y\sqrt{\alpha}}{\sqrt{\alpha}+1}\right)^l \frac{1}{l!} \sum_{p=0}^{\infty} \frac{\left(-\frac{\phi}{k}-M\right)^p \Gamma\left(\frac{l}{2}+1\right)}{p!\,\Gamma\left(\frac{l}{2}-p+1\right)s^{\zeta p - \frac{\zeta l}{2}}} - G_r \sum_{l=0}^{\infty}\left(-y\sqrt{\frac{p_r}{1+N}}\right)^l \frac{1}{l!}$$

$$\times \sum_{p=0}^{\infty}\left(\frac{\alpha(N+1)}{p_r(\alpha+1)}\right)^{p+1} \sum_{q=0}^{\infty} \frac{\left(-\frac{\phi}{k}-M\right)^q \Gamma(p+1)}{q!\,\Gamma(p-q+1)s^{2+\zeta-\frac{l\zeta}{2}+q\zeta}} - G_m \sum_{l=0}^{\infty} \frac{(-y\sqrt{S_c})^l}{l!}\sum_{p=0}^{\infty}\left(\frac{\alpha}{S_c(\alpha+1)}\right)^{p+1}$$

$$\times \sum_{q=0}^{\infty} \frac{\left(-\frac{\phi}{k}-M\right)^q \Gamma(p+1)}{q!\,\Gamma(p-q+1)s^{\zeta-\frac{l\zeta}{2}+q\zeta+1}}, \tag{22}$$

using inverse Laplace transform with convolution property, we get

$$V(y,t) = UH(t)\cos(\omega t) + UH(t)\int_0^t \cos\omega(t-\tau) \sum_{l=1}^{\infty}\left(\frac{-y\sqrt{\alpha}}{\sqrt{\alpha}+1}\right)^l \frac{1}{l!}\sum_{p=0}^{\infty} \frac{\left(-\frac{\phi}{k}-M\right)^p \Gamma\left(\frac{l}{2}+1\right)t^{\zeta p - \frac{l\zeta}{2}-1}}{p!\,\Gamma\left(\frac{l}{2}-p+1\right)}d\tau$$

$$-G_r \sum_{l=0}^{\infty} \left(-y\sqrt{\frac{p_r}{1+N}}\right)^l \frac{1}{l!} \sum_{k=0}^{\infty} \left(\frac{\alpha(N+1)}{p_r(\alpha+1)}\right)^{p+1} \sum_{q=0}^{\infty} \frac{\left(-\frac{\phi}{k}-M\right)^q \Gamma(p+1) t^{\zeta+q\zeta-\frac{l\zeta}{2}+1}}{q!\,\Gamma(p-q+1)}$$

$$-G_m \sum_{l=0}^{\infty} \frac{(-y\sqrt{S_c})^l}{l!} \sum_{p=0}^{\infty} \left(\frac{\alpha}{S_c(\alpha+1)}\right)^{p+1} \sum_{q=0}^{\infty} \frac{\left(-\frac{\phi}{k}-M\right)^q \Gamma(p+1) t^{\zeta+q\zeta-\frac{l\zeta}{2}}}{q!\,\Gamma(p-q+1)}, \quad (23)$$

implementing generalized Mittage-Leffler function on equation (23), we obtain the compact form of velocity field

$$V(y,t) = UH(t)\cos(\omega t) + UH(t)\int_0^t \cos\omega(t-\tau) \sum_{l=1}^{\infty} \left(\frac{-y\sqrt{\alpha}}{\sqrt{\alpha}+1}\right)^l \frac{1}{l!} M_{\zeta,-\frac{\zeta}{2}}^{\frac{l}{2}-p+1}\left(-t^\zeta \frac{\phi}{K} - t^\zeta M\right) d\tau$$

$$-G_r \sum_{l=0}^{\infty} \left(-y\sqrt{\frac{p_r}{1+N}}\right)^l \frac{1}{l!} \sum_{p=0}^{\infty} \left(\frac{\alpha(N+1)}{p_r(\alpha+1)}\right)^{p+1} M_{\zeta,-\frac{l\zeta}{2}+\zeta+2}^{p-q+1}\left(-t^\zeta \frac{\phi}{K} - t^\zeta M\right)$$

$$-G_m \sum_{l=0}^{\infty} \frac{(-y\sqrt{S_c})^l}{l!} \sum_{p=0}^{\infty} \left(\frac{\alpha}{S_c(\alpha+1)}\right)^{p+1} M_{\zeta,-\frac{l\zeta}{2}+\zeta+1}^{p-q+1}\left(-t^\zeta \frac{\phi}{K} - t^\zeta M\right), \quad (24)$$

Where, the property of generalized Mittage-Leffler function is

$$t^{\eta-1} E_{\xi,\eta}^{\chi}(Q) = t^{\eta-1} \sum_{j=0}^{\infty} \frac{(Q)^j \Gamma(\chi+j)}{\Gamma(\chi)\Gamma(\xi j+\eta)} = M_{\xi,\eta}^{\chi}(Q), \quad Re(\xi) > 0, \quad Re(\eta) > 0. \quad (25)$$

Employing identical procedure we have solution for Case-II: $V(0,t) = UH(t)\sin(\omega t)$,

$$V(y,t) = UH(t)\sin(\omega t) + UH(t)\int_0^t \sin\omega(t-\tau) \sum_{l=1}^{\infty} \left(\frac{-y\sqrt{\alpha}}{\sqrt{\alpha}+1}\right)^l \frac{1}{l!} M_{\zeta,-\frac{\zeta}{2}}^{\frac{l}{2}-p+1}\left(-t^\zeta \frac{\phi}{K} - t^\zeta M\right) d\tau$$

$$-G_r \sum_{l=0}^{\infty} \left(-y\sqrt{\frac{p_r}{1+N}}\right)^l \frac{1}{l!} \sum_{p=0}^{\infty} \left(\frac{\alpha(N+1)}{p_r(\alpha+1)}\right)^{p+1} M_{\zeta,-\frac{l\zeta}{2}+\zeta+2}^{p-q+1}\left(-t^\zeta \frac{\phi}{K} - t^\zeta M\right)$$

$$-G_m \sum_{l=0}^{\infty} \frac{(-y\sqrt{S_c})^l}{l!} \sum_{p=0}^{\infty} \left(\frac{\alpha}{S_c(\alpha+1)}\right)^{p+1} M_{\zeta,-\frac{l\zeta}{2}+\zeta+1}^{p-q+1}\left(-t^\zeta \frac{\phi}{K} - t^\zeta M\right). \quad (24)$$

## 6. Limiting Cases

### 6.1 Ordinary Solution for Temperature Distribution When $\zeta \to 1$.

Letting $\zeta \to 1$ in equation (11) and simplifying with the help of convolution theorem and fact of fractional calculus, we get

$$T(y,s) = \int_0^t (t-\tau) \frac{\left(-y\sqrt{\frac{p_r}{(1+N)}}\right)}{s\sqrt{\pi t^3}} e^{-\frac{\left(y^2 p_r/(1+N)\right)}{4t}} d\tau, \quad (25)$$

## 6.2 Temperature Distribution in the Absence of Thermal Radiation When $N \to 0$.

Taking $N \to 0$ in equation (11), we recovered the solution of temperature distribution without thermal radiation in terms of Robotnov-Hartley function as

$$T(y,s) = \int_0^t (t-\tau) \, \boldsymbol{F}_{-\frac{k}{2}}\left(-y\sqrt{p_r},t\right) d\tau, \tag{26}$$

Where, the property of Robotnov-Hartley function is

$$\boldsymbol{F}_\alpha(\beta,t) = \sum_{i=0}^{\infty} \frac{\beta^i \, t^{Di-1}}{\Gamma(Di)}, \tag{27}$$

## 6.3 Ordinary Solution for Mass Concentration When $\zeta \to 1$.

Substituting $\zeta \to 1$ in equation (15) and simplifying by using the fact of fractional calculus for error complementary function, we get

$$T(y,s) = erfc\left(\frac{y\sqrt{S_c}}{2\sqrt{t}}\right), \tag{28}$$

It is also pointed out that we can retrieve various solutions for velocity field for instance, taking $\zeta \to 1, M \to 0$, and $\emptyset \to 0$ in equation (25) solutions can be recovered for ordinary differential equations, without magnetic effects and without porous medium respectively.

## 7. Concluding Remarks

This Portion is dedicated to highlight the major impacts of heat and mass transfer in magnetohydrodynamic casson fluid embedded in porous medium. The generalized solutions have been traced out for the temperature distribution, mass concentration and velocity profiles under the existence and non-existence of transverse magnetic field, permeability and porosity. The corresponding solutions of temperature distribution and mass concentration, velocity profiles have been expressed in terms of newly defined generalized Robotnov-Hartley function, wright function and Mittage-Leffler function respectively. All the corresponding solutions fulfill necessary conditions (initial, natural and boundary conditions) as well. Caputo Fractionalized solutions have been converted for ordinary solutions by substituting $\zeta = 1$. Some similar solutions for the temperature distribution, mass concentration and velocity profiles have been particularized form generalized solutions as the limiting cases. In order to emphasize vivid effects of implemented rheology, we have depicted various graphs listed as 1-7. However, we perused main finding enumerated as under:

(i). Fig. 1 is prepared to highlight the effects of varying time, it is noted that scattering behavior of fluid flow is perceived in creeping for mass concentration, temperature distribution and velocity field at whole domain of heated plate.

(ii). The characteristics of fluid flows with increasing fractional parameter ($\zeta$) are depicted in Fig. 2. It is noted that velocity field and temperature distribution are increasing with increment in fractional parameter ($\zeta$) in the range $0.2 \leq \zeta \leq 0.6$. Further as predicted, mass concentration has strong influence on fractional parameter ($\zeta$).

(iii). Fig. 3 shows the impacts of prandtl number ($P_r$) on mass concentration. Prandtl number ($P_r$) has vital role on the process of mass transfer. It is seen in Fig.3 for mass concentration that ratio of thickness between concentration boundary layer and viscous is characterized. In continuation, identical effects between prandtl number ($P_r$) and schmidt number ($S_c$) are seen in temperature and concentration. Temperature distribution scattered significantly while increasing thermal radiation $N$.

(iv). While increasing thermal Grashof number, mass Grashof number, transverse magnetic field and porosity, similar impacts have been observed in opposite direction over the boundary for velocity field in Fig. 4 and 5.

(v). Fig. 6 displays comparison between four models of fluid namely (i) fractionalized Casson fluid with porous, (ii) fractionalized Casson fluid with transverse magnetic field, (iii) fractionalized Casson fluid without porous and (iv) fractionalized Casson fluid without transverse magnetic field. Comparison of four models has been underlined at different time for velocity field. It is observed that among four models of fluid, fractionalized Casson fluid without transverse magnetic field moves fastest at all the time as expected. This is due to the fact that absence of effective rheology (magnetized material, chemical reaction, porous, permeability, etc.) on fluid.

(vi). On contrary, Fig. 7 is depicted for ordinary casson fluid ($\zeta = 1$) with four models as discussed in (v). It is observed that contrasting behavior of fluid is traced out in comparison with Fig. 6. The same comparison can be made on mass concentration and temperature distribution.

**Recommendations of Future Direction**

It is very significant to high light few limitations regarding this research work. Such limitations will not simply support researchers to analyze this work but also provide extension of this research work. However, following assumptions and limitations can be considered

- To analyze the same problem by employing newly defined Caputo-Fabrizio fractional derivative.
- Same problem can be evaluated under the consideration of Newtonian heating.
- To extend present results using slip condition/assumption for accelerating and oscillating heated plate.

**Acknowledgement**

The author Kashif Ali Abro is highly thankful and grateful to Mehran university of Engineering and Technology, Jamshoro, Pakistan for generous support and facilities of this research work.

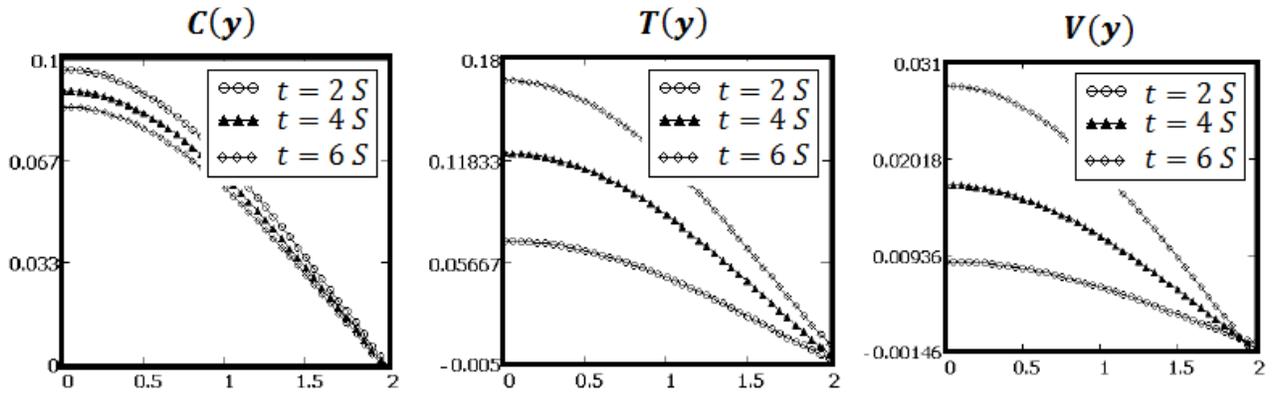
Fig. 01. Profile of mass concentration, temperature distribution and velocity field.

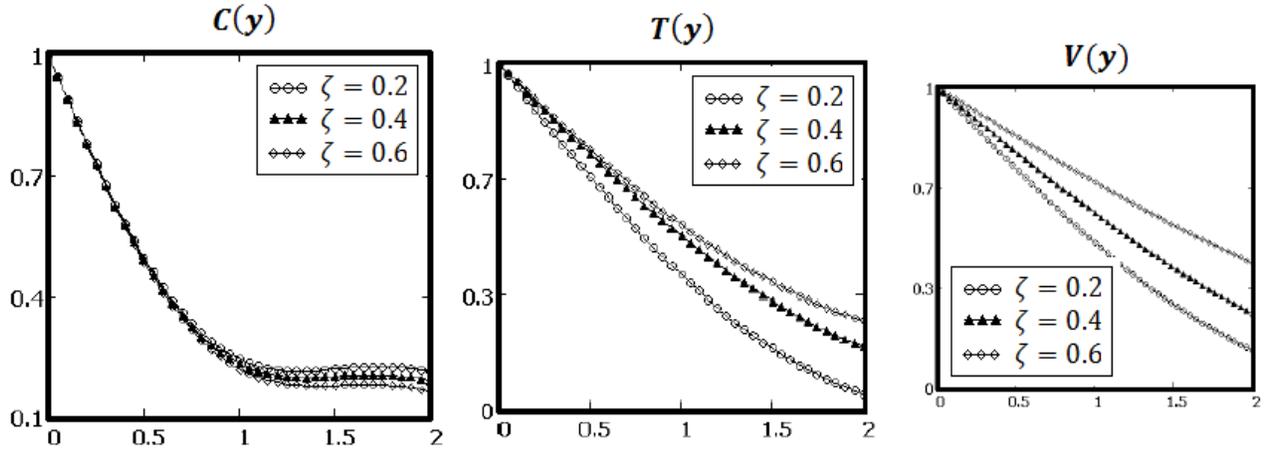
Fig. 02. Profile of mass concentration, temperature distribution and velocity field.

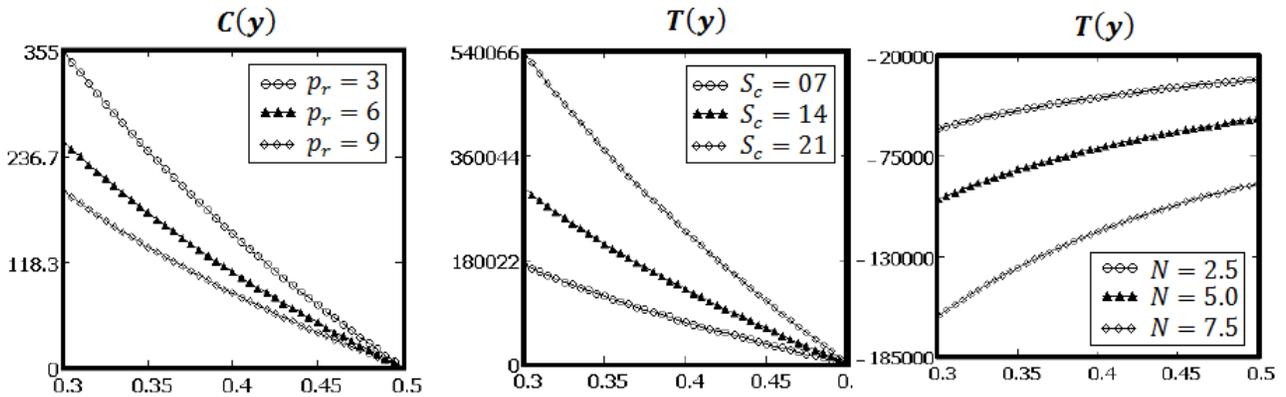
Fig. 03. Profile of mass concentration and temperature distribution.

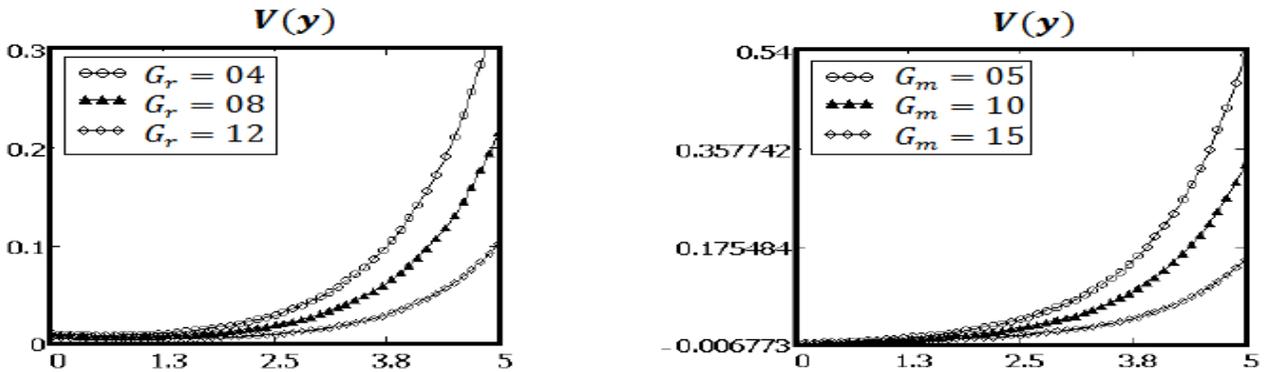
Fig. 04. Profile of velocity field.

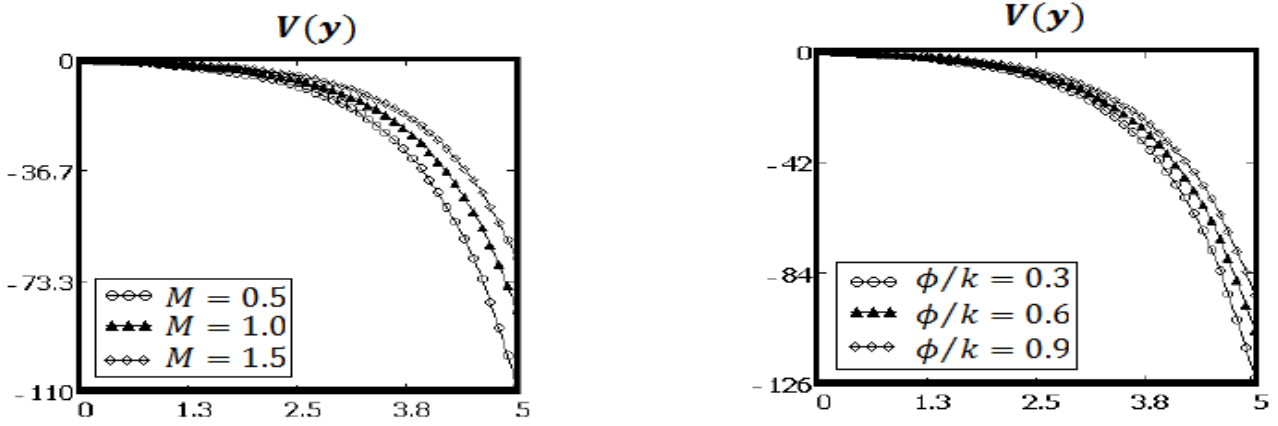
Fig. 5. Profile of velocity field.

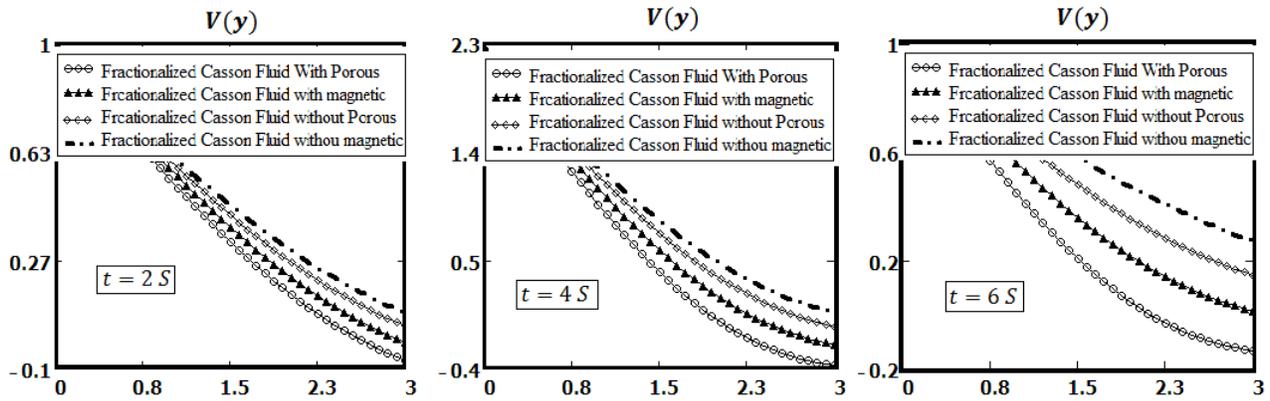
F.g. 06. Profile of velocity field.

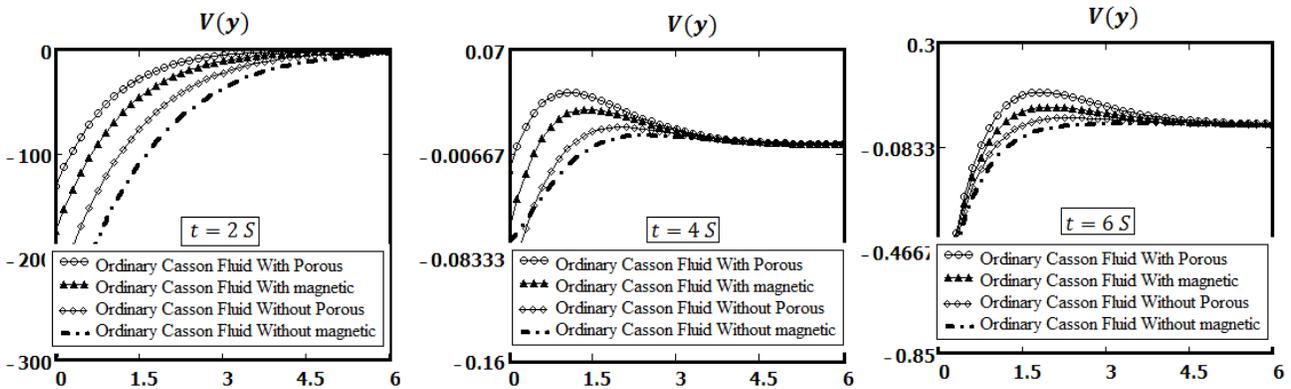
Fig. 07. Profile of velocity field.